# Using physiological measures in conjunction with other usability approaches for better understanding of the player's gameplay experiences

**Pejman Mirza-Babaei**

**Graham McAllister**

Interact Lab, University of Sussex, UK

*The goal of video games is to challenge and entertain the players. Successful video games deliver experience that impact players on a level of arousal. Therefore undertaking a user experience (UX) study is crucial to ensure that a game achieves both critical and financial success. However, traditional usability methods (observation, subjective reporting, questionnaire, and interview) have a number of limitations on game user research.*

*In this study we capture player's physiological measures during a gameplay session, to indicate micro-events that have caused changes in their body signals. At the post-gameplay interviews we ask participants to comment and describe their feelings on the selected events. The aim of this study is not to over-interpret physiological measures, but on using blips in measures to help identify key points in a game, which we then use to investigate further with the participant.*

*This approach provides a method that can identify not only the negative user experience and usability issues but also the events which have a positive impact on player's experience.*

## INTRODUCTION

Within the rapidly growing computer and console game market, researchers are using emerging technologies to develop enhanced play environments. The wide variety of video games make them a popular type of entertainment for a broad range of consumer groups. PriceWaterhouseCoopers recently stated that video game industry is expected to rise from $41.9 Bn of global sales in 2007 to $68.3 Bn in 2012, a compound annual growth rate of 10.3% (Bond, 2008).

While film and game industries are enjoying a massive technology improvement in the last few years (such as James Cameron's *Avatar*), researchers and developers still suffer from a lack of effective evaluation methods.

Although Human Computer Interaction (HCI) methods have made progress in understanding product usability, applying the current methods to identify user experience issues is still a challenge for UX researchers.

Our research interest is in how to combine physiological metrics with other usability methods to identify user experience issues for gameplay environments. This paper explains why we need such an approach and suggests a method to adapt it.

**Traditional playtest methodologies**

Current methods of evaluating entertainment technologies include both subjective and objective techniques. The most common methods are subjective self-reports through questionnaires, interviews, and focus groups (Fulton & Medlock, 2003) as well as objective reports through observational video analysis (Lazzaro, 2004).

**'*Focus Groups'*** are ideal at the start of projects for swiftly getting a sample of player's opinions and their general feeling about the project. They are useful for determining what players expect, need and desire. The open discussion with a group of players is good for testing and getting feedback on whether the idea behind a game makes sense, which features would players expect or other similar questions that designers need to know early on when embarking upon a new projects. However it can be difficult to include all the participants in the discussion equally. 'Focus Groups' can even run late on in the production process to do a reality check and fine-tune the message (Krug, 2000).

*'Observation'* involves watching the player interact with the game; we observe how they act, by monitoring their behaviour. The observer can be in the same room, or watch them remotely. Though direct observation and artificial environment can produce biased results. Observation sessions are easy to setup and run. However, understanding player behaviour requires precise interpretation and unless the video data is captured, some important events can be missed.

Using 'observation' is a rich source of data, but studying observational data as an indication of human experience is a lengthy and difficult process that needs to be undertaken with great care to avoid biasing the result (Marshall & Rossman, 1999). In HCI, behavioural observation logs are common analysis tool. They can provide a basis for a detailed analysis of usability (Pagulayan et al., 2003), fun and game experience (Poels et al., 2007).

*'Think aloud'* or *'verbal reports'* involve asking the players to talk out loud and describe their actions, feelings and motivation while they are playing the game. The aim is to get inside their thinking processes 'in the moment'. It can reveal previously unnoticed details and can give immediate feedback to the action. However, it is unprompted and may be unnatural for the participants, which can affect the gameplay experience. Also if the timing aspect of the game is integral to the game mechanic, then getting the player to talk will affect this.

'Think aloud' techniques cannot effectively be used within game testing sessions because of the disturbance to the player and ultimately the impact they have on game play (Nielsen, 1992).

*'Interviews'* and *'questionnaires'* are the basis for user-feedback gathering, where the player is asked a series of structured questions. These methods can address specific issues, but sometimes these issues remain unsolved as there is always the potential for biased questions and answers. Also often people don't remember exactly what the motives for their actions were. By recording the game video, it is possible to take note of moments of interest during the test, then replay those sections and ask the player what they were feeling in order to facilitate recall.

Subjective reporting through 'questionnaires' and 'interviews' are generalisable, convenient, and amenable to rapid statistical analysis. Yet, they only generate data when a question is asked, and interrupting gameplay to ask a question is too disruptive (Mandryk et al., 2006).

Explained traditional methods have been adopted with some success for evaluating entertainment technologies. However, the success of a play environment is determined by the process of playing, not the outcome of playing (Pagulayan et al., 2003). We must consider this when evaluating user experience (UX) on a video game. Although traditional methods can identify major gameplay navigation and content issues, as well understand the attitudes of the users; they suffer from low evaluation bandwidth, providing information on the finished experience, rather than continuously throughout the course of the game (Mandryk & Atkins, 2007). They can interfere with game-play and create an artificial experience, producing inaccurate results.

**Biometric method**

Cutting edge technologies enable UX experts to use physiological measurements for testing or quantifying player's feeling. Biometrics is the science of capturing and analysing signals directly from the player's body. They can show how different player's body react to the events on-screen. Galvanic Skin Response (GSR), Heart Rate (HR), Electroencephalography (EEG), and Electromyography (EMG) are the most common metrics systems in use for game research. Since they measure biological responses, they are instinctive and cannot be falsified. They can help reveal a player's experience and physiological state.

Psychologists use physiological measures to differentiate between human emotions such as anger, grief and sadness (Ekman et al., 1983). Physiological metrics have recently been used in Human computer Interaction. Some researchers have used GSR and cardiovascular measures to examine user response to well and ill-designed web pages (Ward & Marsden, 2003). More recently, some researchers in UX have used physiological measurements to evaluate emotional experience in play environments. For example, Mandryk has used HR, GSR and EMG to create a modelled emotion for interactive play environment (Mandryk et al., 2006) and she has examined physiological responses to different interactive play environment (Mandryk & Inkpen, 2004). Nacke has created a real-time emotional profile (flow and immersion) of gameplay by using facial EMG and GSR (Nacke & Lindley, 2008).

Although using a response profile for a set of physiological variables enables scientists to go into more details with their analysis and allows for a better correlation between response profile and psychological event (Cacioppo et al., 2007). But, changes in the physiological signals can be responses to external activity or can be in anticipation of something not otherwise observed. Moreover, specific types of measurement of different responses (such as GSR, EMG, ECG and EEG) are not trustworthy signs of well-characterised feeling (Cacioppo, 2007). The often described 'many-to-one' relation between psychological processing and physiological response (Cacioppo et al., 2007) allows for physiological measures to be linked to a number of psychological structures (for example; attention, emotion, information processing) (Nacke & Lindley, 2008). Ambinder states; "Some responses or measurements are difficult to correlate with something specific that happened in the game" (Onyett, 2009).

This paper describes a study that is cautious not to over-interpret physiological measures, but on using glitch in measures to identify micro-events in a game that we want to investigate further with participants. In order to select micro-events, we monitor players' faces, biometric data, verbal comments and video game output of a gameplay session. The main goal is to establish and validate a method that can specify the key moments in a game that impact players on arousing level. Implementing this method in conjunction with other usability methods can help us to reduce the impact of the aforementioned limitations in traditional UX techniques. Moreover, this approach can emphasize not only user experience and usability problems but also moments that have a positive impact upon the players' feelings.

In order to validate this method, we conducted an experiment in February 2010 in our dedicated game user research laboratory (Vertical Slice) at the University of Sussex in the United Kingdom.

In the following paragraphs we give an overview of our experimental methodology; and how we report the findings and results. And finally there is a conclusion as well as a discussion and a detailed prognosis for future work.

## METHOD

Participants played the first two levels of '*Call of Duty: Modern Warfare 2 (MW2)*' (Call of Duty: Modern Warfare 2, 2009) and '*Haze*' (Haze, 2008) both on the default setting and on the Sony PlayStation 3 platform. Metacritic (Metacritic, n.d.) review scores 94 of 100 for *Call of Duty: Modern Warfare 2*. This means this game is highly accessible, meets expectations, has good tutorials and positive player feedback. On the other hand, *Haze* metascore is 55 of 100 (Metacritic, n.d.), which means the game has key usability issues that impacts enjoyment. These games have been selected specifically to show how the players' bodies react to a well designed and a poorly designed game. These two different conditions assess the independent variables of game experience.

Data was collected from six male participants, aged 20 to 31. They were all higher education students at the University of Sussex. Before the experiment began all participants filled out a background questionnaire, which was used to gather information on their experiences with computer games, game preference, console exposure and personal statistics such as age. Participants were recruited carefully; none of them had played *Call of Duty: Modern Warfare 2* or *Haze* before. Participants were casual gamers, playing either computer games or console game frequently. All the participants owned a personal computer and they had played on Sony PlayStation 2 or 3 platforms before. All the participants played game at least twice a week. Half of them preferred to play alone and the other half preferred multiplayer modes. All of them started to play digital games when they were younger than 11 years old. It is important to note that none of the subjects received any compensation for their participation in the experiment.

Half of the participants were asked to play the first and second levels of *Call of Duty: Modern Warfare 2* in the normal difficulty mode. The other half played the first and second levels of *Haze* with the same difficulty settings.

All experiments were conducted between 10:00am and 7:00 pm, with each experiment session lasting approximately 2 hours. Upon arriving, after a brief description of the experiment procedure, participants signed a consent form. They were then fitted with physiological sensors. The gameplay session took around 75 minutes, depending on how fast they finished two levels.

The post-gameplay interview was conducted soon after they had finished the gameplay session. From the earlier studies we learnt that it is better to conduct the interview as soon as possible so that the participants can remember most of their actions and thoughts. The interview was based on the selected events from the gameplay session, namely those events selected by monitoring changes in participants' physiological measures, their verbal report, sitting position, facial expressions and the game output. With the participants, we looked at selected moments of their gameplay video and they described their feelings on those moments but most importantly *'why they felt that way'*.

Physiological data were gathered using BIOPAC hardware system, sensors and software from BIOPAC System Inc. Based on previous literature, we chose to collect galvanic skin response (GSR) and electrocardiography (EKG). Heart rate (HR) was computed from the EKG signal.

**Galvanic skin response (GSR)**

Arousal is commonly measured using galvanic skin response, also known as skin conductance (Lang et al., 1993). The conductance of the skin is directly related to the production of sweat in the eccrine sweat glands. In fact, subjects do not have to even be sweating to see difference in GSR because the eccrine sweat glands act as variable resistors on the surface. As sweat rises in a particular gland, the resistance of that gland decreases even thought the sweat may not reach the surface of the skin (Stern et al., 2001). Galvanic skin response has a linear correlate to arousal (Lang, 1995) and reflects both emotional responses as well as cognitive activity (Boucsein, 1992).

We measured the impedance of the skin by using two passive SS3LA BAIOPACK electrodes (at 60 Hz). The electrode pellets were filled with TD-246 skin conductance electrode gel and attached to the ring and little fingers of the participant's left hand.

**Cardiovascular measures**

The cardiovascular system includes the organs that regulate blood flow through the body. Measures of cardiovascular activity include heart rate (HR), interbeat interval (IBI), heart rate variability (HRV), blood pressure (BP), and BVP. Electrocardiograms (EKG) measure electrical activity of the heart, and HR, IBI, and HRV can be computed from EKG. HR reflects emotional activity. It has been used to distinguish between positive and negative emotions (Papillo & Shapiro, 1990).

In this study we monitored participant's HR, which was computed from their EKG. To collect EKG we placed three pre-gelled leads with BIOPACK SS2L surface electrodes (at 50Hz) in the standard configuration of two electrodes just above the ankles and one electrode on the right wrist over the veins.

**Playroom**

The experiment was conducted in our dedicated gameplay laboratory (Vertical Slice) at the Sussex University. Our playroom equipped with a Sony PlayStation 3, a Sony 40" flat screen TV, a Sennheiser wireless microphone, a Sony Handycam video camera to capture the player's face and a BIOPAC system to capture physiological data. Participants were seated on a comfortable sofa positioned approximately one meter from the TV and the camera. The playroom specifically is designed and decorated to simulate an actual living room in order to reduce the impact of artificial experience. The game footage, the camera recording the player's face, and the screen containing the physiological data (GSR and HR) were synchronized into a single screen. This screen is digitally recorded and displayed on another screen in an isolated observation room, where we were observing and controlling the gameplay session. The digital recording also contained audio of the participant's comments and the game audio output from an attached microphone.

## RESULTS

The results listed below validated the effectiveness of this approach to understand players' thoughts and behaviour in the games. The results were compared to identify which types of game events affect the participant's gameplay experiences.

*Repeating moments*

HR and GSR signals increase at the beginning of each new level in both games. Participants all commented that it was because of the excitement of a new mission and getting ready to play. Similar peaks were observed at the end of each level. All the players explained that it was due to the happiness and enjoyment of completing the mission and getting ready for the next one.

Repeated increases in players' GSRs signal were noticed when they died and got back into the game, which explained as: "knowing what is going to happen and the anticipating that". In the same way, when a participant died many times in a scene: "it was frustrating, I couldn't figure out what to do". In one event, a participant commented about the increase in his GSR signal when he died, as the respawn location was considerably far behind the point of death, meaning that a lot of progress was lost.

At the beginning of the first level of *Haze*, increase in player's GSR signal occurred after he walked in different directions in the jungle for about two minutes. During the post-gameplay interview he said: "I was feeling lost and not in control of the game".

*Feedback and direction*

During different events in *Haze*, players described most of the changes in their signals, as they were not sure if they were doing a right thing. For example Lev (participant 2) expressed: "I was not sure if I could still drive my buggy or if it was broken. I've started driving it again, but was not sure if it was going to explode soon or not. Eventually, it did".

When we asked him about a change in his signals at the point he got some instructions on screen, Lev continued: "in that event, when I was driving, there were four lines of text, with a small font on the screen. I couldn't read them and was worried they might be important instructions or directions".

In another event he described a change in his GSR signal when he was shot to death as: "there were little arrows on the screen telling me from which direction the enemies are shooting at me, but no one was there. It was so confusing. I couldn't see who was shooting at me".

Bob's (participant 1) GSR and HR signals decreased when he got off the buggy after driving it up a hill. Later on in the interview he commented, "I was not sure what to do, there were no clear instructions; should I walk or continue driving? I decided to walk but was worried that it was not the correct decision". (Figure 1)

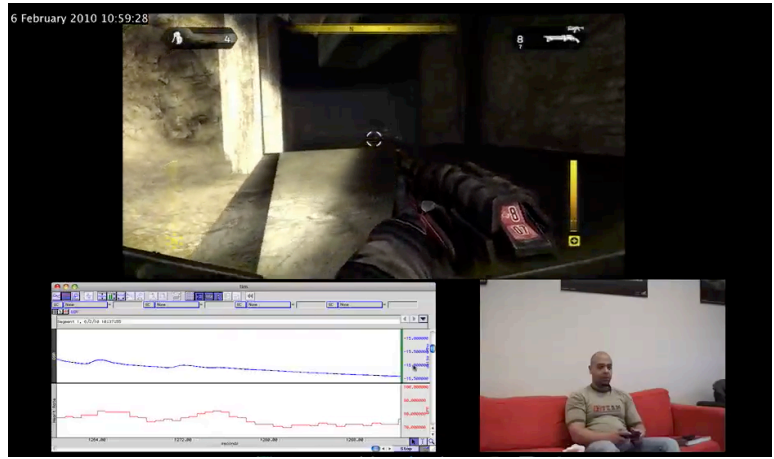

**Figure 1: Player's GSR and HR when he got off the buggy**

*Cutscenes*

In both games, repetitive change in players' GSR signal occurred while they were watching cutscenes. In the following there are some of the comments players mentioned in the post-gameplay interview about the changes in their physiological responses.

We noticed an increase in Bob's GSR signal while he was watching a cutscene in the middle of the first level of *Haze*; we asked him if he was enjoying this clip? In the post-gameplay interview Bob commented that "this was very boring and I couldn't skip it". Another participant describes the change in his signal at the same clip as: "I was not sure if I was walking in a right direction. I was lost in the jungle, so when the clip started I was happy because I realised I was in the correct location". Similar changes in GSR signal were indicated with other participants while they were watching this clip or other cutscenes in this game. They similarly commented: "I wanted to be able to skip cutscenes unless they give orders or directions…these clips are boring especially for an FPS game (First Person Shooter)…while we were on helicopter at the beginning of the second level, I was expecting to receive the next mission briefing".

On the other hand, in *MW2*, the player comments: "the cutscenes are interesting, especially because they are a briefing of the next level and it is great that I can skip watching it…they are replaced the loading screen…cutscenes during the levels are extremely short". During one of the clips a player's GSR signal changed and he commented: "I like the Indian Ocean and it was zooming into it".

*Weapons*

Frequent changes occurred in GSR and HR when players recieved their favourite guns.

In *Haze*, we observed an increase in Bob's GSR signal when he was using the sniper rifle, our question were if he liked the weapon? In the post-gameplay interview he mentioned: "In that event, I killed four enemies all together; all of them came to the game scenario from the same location, and it was very easy to kill them all. I was expecting a tougher experience for an FPS game". (Figure 2)

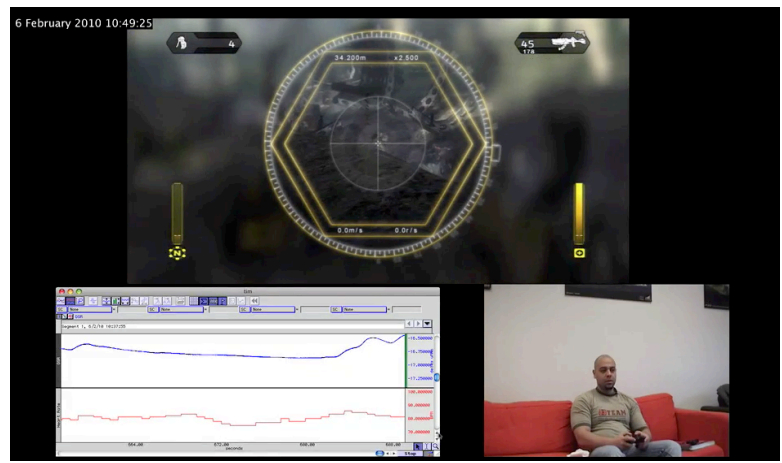

Figure 2: Increase in Player's GSR when he was using the sniper rifle

There were several increases in GSR signals when players tried to use sniper rifles in different events in *Haze*. Similarly, all of their comments were: "It is always cool to snipe in an FPS game, but in this game aiming was so difficult and it was hard to kill enemies." One participant commented: "enemies were moving too much".

A change in Bob's GSR signal was identified while he was trying to kill an enemy with shotgun. Later in the interview he mentioned: "I was shooting at him with shotgun repeatedly, but he didn't die. Shotguns are usually extremely powerful weapons and it should only take one or two shots to kill someone".

In *MW2* a participant commented: "I like the fact that in each level I need to use different guns depending on environmental conditions". The players' GSRs change when they try to snipe and use grenades. An example of this is shown in the comment: "it is always fun to use grenades. In this game they give you an authentic feeling".

During the same game, repeated increases in one of the player's signals were observed before he reloaded his gun. Later in the interview he commented: "I've chosen a machinegun because it is so powerful but it takes time to reload it. It is annoying, but it is also realistic which I like. Every time I wanted to reload it I was afraid the enemy might attack me before the gun was reloaded".

*New features*

In the second level of *Haze* a new feature is introduced into the game (a buggy). Bob's GSR signal increased when he started riding the buggy. Later at the post-gameplay interview he commented: "I was happy seeing the buggy and knew I could drive it and it was going to be fun. It was not easy to get on it. There were four seats available; the driver seat, right or left passenger seats and the top slot armed with a machinegun, I was able to sit in any of the seats, but to continue the game I had to sit as the driver. Also I had to approach the buggy from behind to be able to sit as the driver. It was confusing. Nevertheless, driving it was fun as soon as I got used to controlling it". He continued: "it was quite fun at the beginning, but later on in the level I got bored with it".

Similar changes were observed in other participants' GSR signals, when they wanted to get on the buggy. All of them commented that: "getting on the buggy was complicated…since we have to drive the buggy to continue the game, why can we sit on the other seats?...I wanted to be able to sit as the driver when I approach the buggy from different sides".

Bob's GSR signal was raised when he was engaged with an enemy's car for the first time in *Haze*, we thought it was because it was the first time he saw an enemy vehicle in the game; but at the post-gameplay interview he mentioned: "Since I was driving my buggy, I was expecting to be able to follow and shoot at the enemy's car, but the car went to a closed area like a warehouse and we were both stuck there shooting at each other. It would be much better if I could follow and shoot at it while driving fast".

In *MW2*, at the beginning of the second level players had to climb an ice rock with a pickaxe. It was something new in the game, so we could see a constant increase in all the players' GSR signals, which could mean a high level of excitement. All players commented that event as: "so interesting, so awesome, I haven't seen anything like this before". Later on at the end of the second level, players had to escape the enemies with a snowmobile. The picture below shows how the player body reacted to that event. All the participants noted that they enjoyed it a lot. One participant said: "the best thing was that riding it was not very difficult, since it is not a racing game. I enjoyed it so much, especially when I had to use my gun as well. It was realistic, I even had to reload my gun too". (Figure 3)

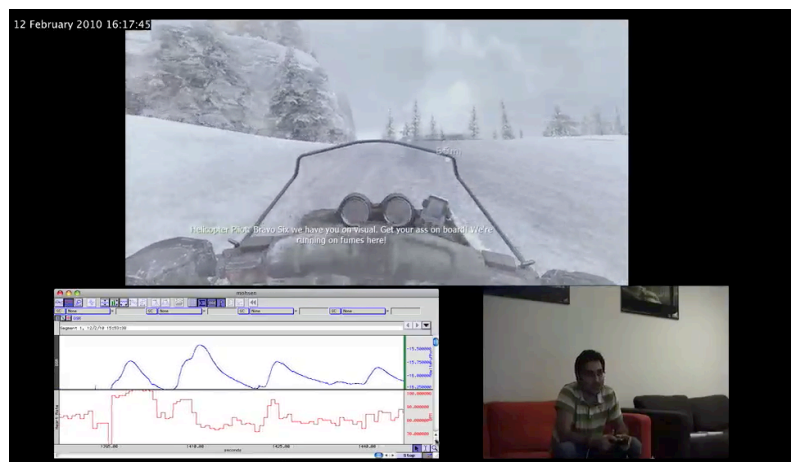

**Figure 3: Changes in player's HR and GSR while riding a snowmobile**

*Playability*

In *Haze*, Bob described the change in his GSR signal when his teammate fired the machinegun on the buggy as: "I saw my teammate shooting at enemies with the machinegun while I was driving, that was cool. I like machineguns and seeing my buddy using it gives me a better sense of security and safety".

Later on at another event he commented on the change as: "The enemies were shooting at me and my buggy, no one was at the machinegun. There was one of my teammate on the buggy but he was not using the machinegun, I was thinking why he did not sit at the machinegun".

And few minutes later when he saw that a new teammate had joined them and sat at the machinegun, his GSR signal changed and he described it in the post-gameplay interview as: "I saw my teammate and I knew he was going to sit at the machinegun, so I felt more secure and in control. Why is the game not designed in a way that my teammates sit at the machinegun straight away? Also, I was wondering if I could fire the machinegun while someone else drove the buggy". (Figure 4)

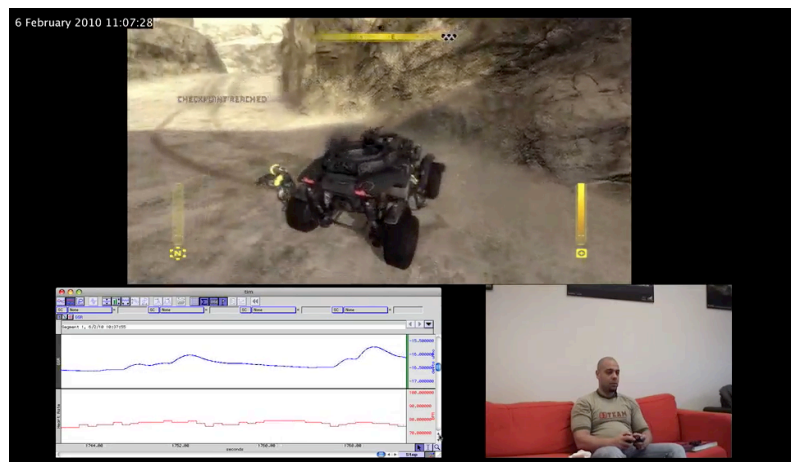

Figure 4: Increase in player's GSR signal when he saw a new teammate

Further changes in Bob's GSR signal occurred while he was using the machinegun to kill an enemy solider. He mentioned earlier that he liked to use the machinegun, so perhaps that is why his GSR signal was changing? During the interview he described his feeling at that moment as: "I was shooting at the enemy, hiding behind an aluminium board, so I thought I must have killed him, since he was hiding behind a very thin board. I was shooting with a powerful gun, yet he was alive".

Bob commented about the increase in his signal when he found a new buggy in the game: "I saw a new buggy, but wasn't sure how I could get there, I needed to jump off a hill. But I died when I tried to jump of a same size hill in the previous level. So I was not sure if I could jump here or not. Since that was the only way to get there I jumped and nothing happened to me. I think that the game rules changed between different levels".

The two screenshots below (Figure 5) compare changes in two players' HR and GSR signals when engaged in a heavy combat:

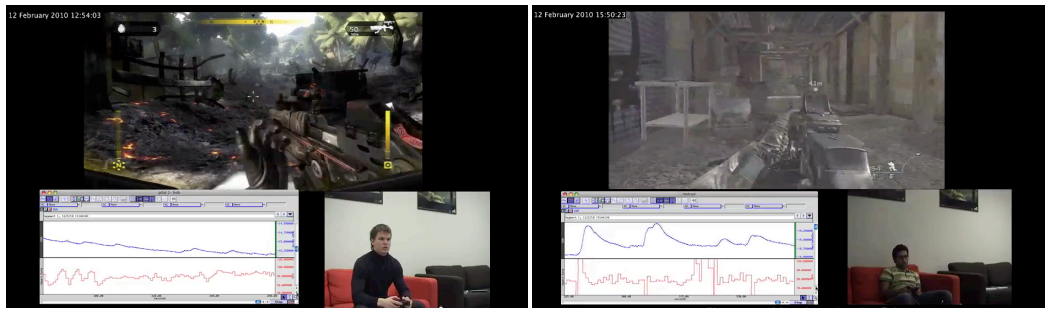

**Figure 5: Players engages in heavy combat in *Haze* (left) *MW2* (right)**

*Game environment*

Lev commented on changes in his GSR signals from the event shown in the below screenshot (Figure 6) from *Haze* as: "falling rocks hit me but nothing happened to me. The game environment does not feel real". In another event he said: "I hit the big container with my little buggy and it moved".

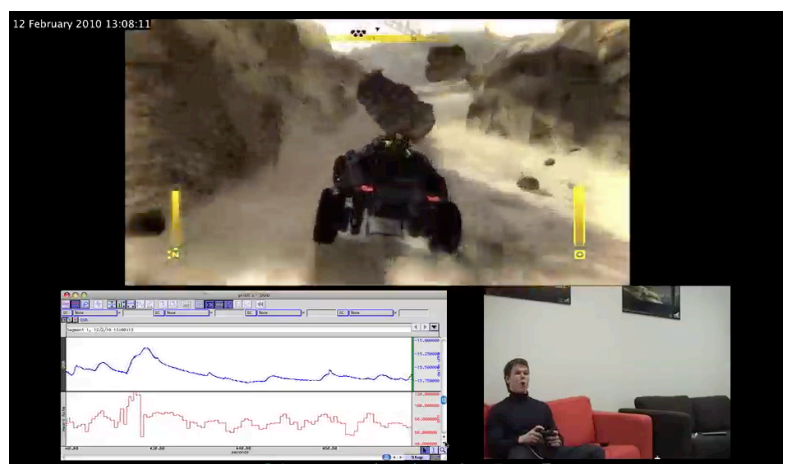

**Figure 6: Changes in player GSR and HR when hit by rockes**

In *MW2* player commented on the increased GSR signals when he entered a building from the street: "You never get tired of this game, the game environment changes frequently, from streets to a building, then back to streets, then on the car".

Another player commented on the change in his signal after a grenade explosion as: "I liked the few second mute after the explosion, it feels like I am not hearing anything. I imagine it is the same in reality".

In the same game, another event the player explained the change in his measurements: "I saw an injured soldier still shooting at me, which was cool…I shot at the tires of a parked car. It was amazing because the tires went flat. This game is so realistic". Similarly from another event: "I love this machinegun, it is very powerful, has a real feeling, yet nothing happened to the trees when I shot at them, I wanted to see changes in the trees".

*Cooling down spot*

In *MW2*, we have noticed similar decreases in GSR and HR for all the players when they finished climbing the ice rock before engaging enemies, at the beginning of the second level. These changes could be assumed as they were getting bored, but at the post-gameplay interview they all commented on that moment as: "a bit relaxing after the heavily taxing moments of climbing the ice, It felt really good and I can't wait to see more of the game". (Figure 7)

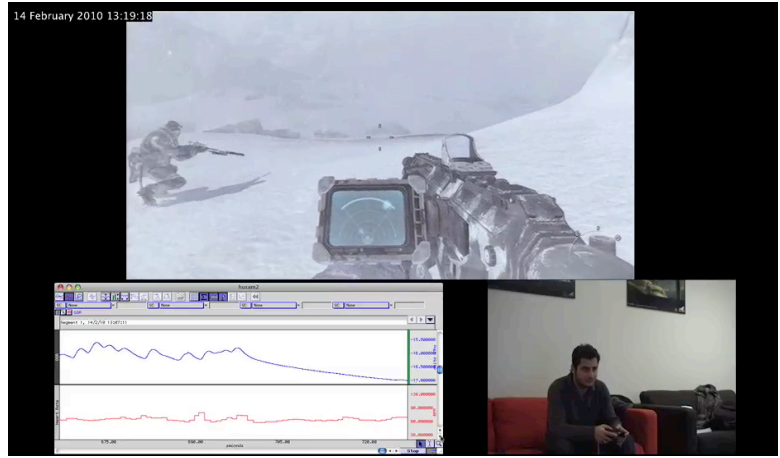

**Figure 7: Decrease in player's GSR after climbing the ice rock**

## CONCLUSION

After analysing the results, we can conclude that the two main benefits of implementing this method in game user research studies are:

1. Sometimes it is too difficult to indicate user experience problems with traditional user research methods, especially if the problem is less obvious. Biometric data are involuntary and objective. By capturing players' body signals during a game-play session, any change in their body reaction can be carefully monitored and that event can then be pointed out for post-gameplay interview session.

2. To confirm findings from other methods or design guidelines. It can also be used as evidence of a problem to prove and back up the found results. This approach can reveal the importance of the impact of usability and playability issues on players' feelings.

Furthermore it can identify critical moments in a game that contribute to a better game-play experience. Biometrics data can indicate if game events are successful in affecting players in arousal level, as well as to specify which scene of a game has more effect on players' feelings.

The results show that implementing this method in combination with other user research methods can help game UX researchers to highlight various usability and user experience issues, but can also point out those moments that have a positive impact upon players' feelings.

The study in this paper was based on comparing two First Person Shooter games, but it is expected that this method can be applied on other genres of games as well. We are currently testing a new car racing game and implementing this method to help us to understand even more about the relationship between game environments and scenarios and the player that is engaging with them.

For this study we used GSR and HR measures, but it is expected that measuring other physiological responses (such as EEG or EMG) could help us to monitor and dissect even more events to discuss with players. However since the sensors need to be attached to the player to collect physiological measures, we have to be cautious with every new type of biometric measurement technique that we may use in the future, for fees of introducing additional factors that may produce biased results.

## ACKNOWLEDGMENTS

Thanks to Vertical Slice and Interact lab colleagues, especially Gareth White and Joel Windels.